\documentclass[prl,twocolumn,
superscriptaddress,
showpacs,
aps
]{revtex4-1}
\usepackage{graphicx}
\usepackage{amsmath}
\usepackage{amsfonts}
\usepackage{amssymb}
\usepackage{hyperref}
\usepackage{color}

\newcommand{\ket}[1]{|{#1}\rangle}
\newcommand{\bra}[1]{\langle{#1}|}

\newcommand{\beq}{\begin{equation}}
\newcommand{\eeq}{\end{equation}}

\usepackage{placeins} 
\usepackage{bm}           

\begin{document}

\title{Dynamical phase transitions to optomechanical superradiance}

\author{Simon B. J\"ager} 
\address{Theoretische Physik, Universit\"at des Saarlandes, D-66123 Saarbr\"ucken, Germany} 

\author{John Cooper}
\address{JILA, National Institute of Standards and Technology and Department of Physics,
	University of Colorado, Boulder, Colorado 80309-0440, USA} 
\address{Center for Theory of Quantum Matter, University of Colorado, Boulder, Colorado 80309, USA} 

\author{Murray J. Holland} 
\address{JILA, National Institute of Standards and Technology and Department of Physics,
	University of Colorado, Boulder, Colorado 80309-0440, USA} 
\address{Center for Theory of Quantum Matter, University of Colorado, Boulder, Colorado 80309, USA} 

\author{Giovanna Morigi} 
\address{Theoretische Physik, Universit\"at des Saarlandes, D-66123 Saarbr\"ucken, Germany} 

\date{\today}

\begin{abstract}
We theoretically analyze superradiant emission of  light from an ultracold gas of bosonic atoms confined in a bad cavity. A metastable dipolar transition of the atoms couples to the cavity field and is incoherently pumped, the mechanical effects of cavity-atom interactions tend to order the atoms in the periodic cavity potential.  By means of a mean-field model we determine the conditions on the cavity parameters and pump rate that lead to the buildup of a stable macroscopic dipole emitting coherent light. We show that this occurs when the superradiant decay rate and the pump rate exceed threshold values of the order of the photon recoil energy. Above these thresholds superradiant emission is accompanied by the formation of stable matter-wave gratings that diffract the emitted photons. Outside of this regime, instead, the optomechanical coupling can give rise to dephasing or chaos, for which the emitted light is respectively incoherent or chaotic. These behaviors exhibit the features of a dynamical phase transitions and emerge from the interplay between global optomechanical interactions, quantum fluctuations, and noise.\end{abstract}

\maketitle

Superradiance describes the collective emission of light by an ensemble of dipoles. It is a quantum interference phenomenon in the emission amplitudes \cite{Dicke:1954,Gross:1982,Kim:2018} and is accompanied by a macroscopic coherence within the ensemble \cite{Dicke:1954,Gross:1982}. In its original formulation, Dicke considered 
$N$ dipoles confined within their resonance wavelength and showed that their spontaneous decay can be enhanced by the factor $N$ \cite{Dicke:1954}.

Quantum interference is typically lost due to fluctuations in the amplitude and in the phase of the dipole-field coupling. These fluctuations can be suppressed by cooling the atomic medium to ultralow temperatures \cite{Inouye:1999,Baumann:2010} and/or by subwavelength localization of the scatterers in an ordered array \cite{Vogel:1985,DeVoe:1996,Clemens:2003,Fernandez:2007,Habibian:2011,Reimann:2015,Begeley:2016,Neuzner:2016}. When, in contrast, the coherence length of the atomic wave function extends over several wavelengths, superradiant scattering of laser light can manifest through the formation of matter-wave gratings \cite{Inouye:1999,Schneble:2003,Piovella:2001,Nagy:2010,Baumann:2010}. In free-space, superradiant gain can be understood as the diffraction of photons from the density grating of the recoiling atoms, which acts as an amplifying medium \cite{Inouye:1999,Schneble:2003}.  Within an optical resonator, these dynamics can give rise to lasing \cite{Bonifacio:1994,Bonifacio:1994:2,Slama:2007,Bux:2013} and be cast in terms of synchronization models \cite{Cube:2004,Slama:2007}. 

\begin{figure}[h!]
	\center \includegraphics[width=0.8\linewidth]{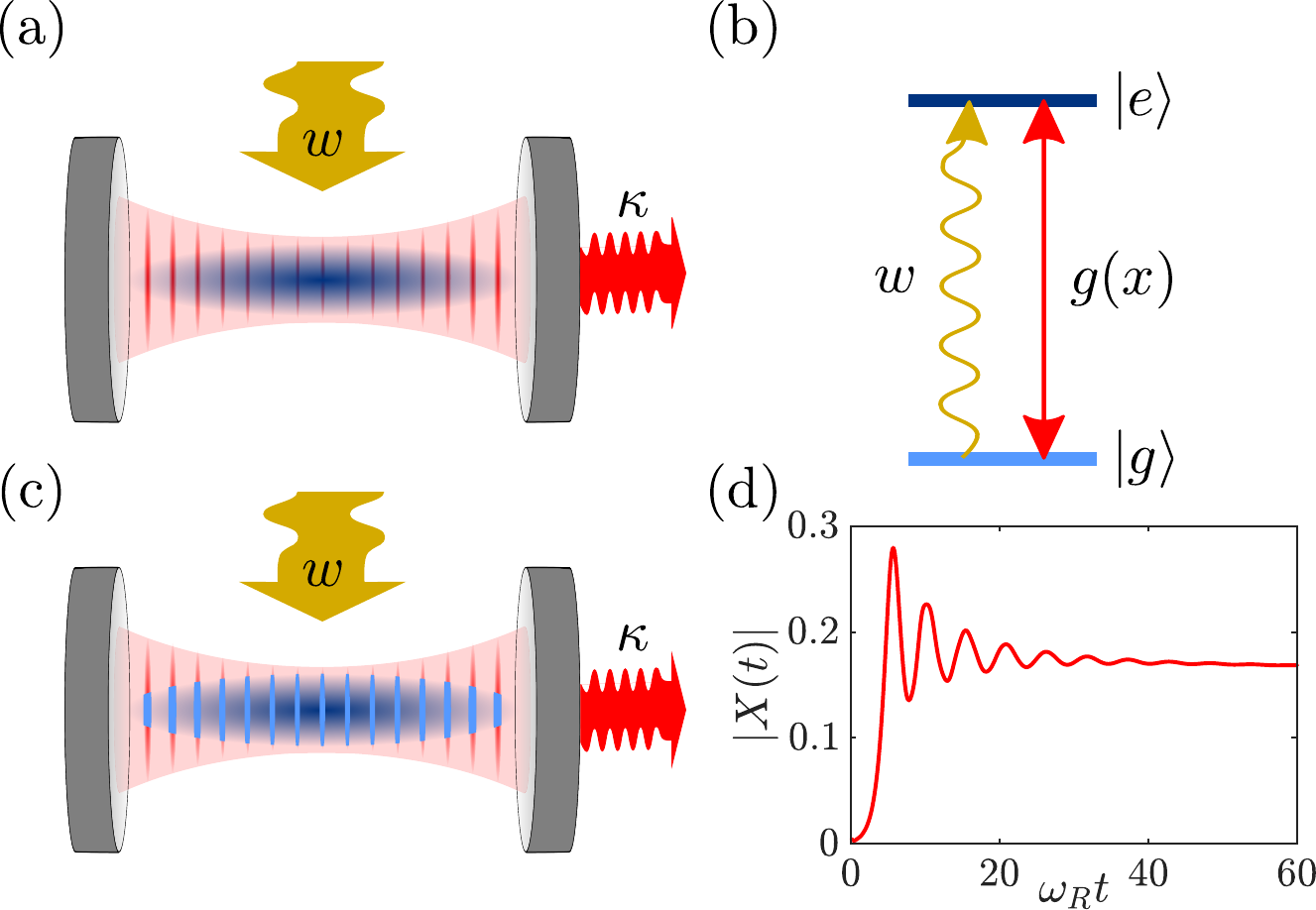}
	\caption{(a) An atomic gas initially forms a Bose-Einstein condensate and is confined within a standing-wave resonator, which emits photons at rate $\kappa$. (b) The metastable atomic transition $\ket{g}\to\ket{e}$ couples to the cavity mode and is incoherently pumped at rate $w$. After the first superradiant decay (c) the atoms form density gratings. (d) The emitted field  $X(t)$ (here in the reference frame of the atomic frequency) becomes coherent for sufficiently large values of $w$, such that one grating is mechanically stable. \label{Fig:1}}
\end{figure}

In this Letter we analyze the interplay between superradiant emission and quantum fluctuations due to the recoiling atoms, when the atoms' dipolar transitions couple to the mode of a lossy standing-wave resonator. In contrast to Refs.~\cite{Inouye:1999,Schneble:2003,Piovella:2001,Nagy:2010,Baumann:2010}, here the atoms are incoherently pumped, as shown in Fig.~\ref{Fig:1}, and therefore no coherence is established by the process pumping energy into the system. The system parameters are in the regime where stationary superradiant emission (SSR) is predicted \cite{Meiser:2009,Meiser:2010:1,Meiser:2010:2,Bohnet:2012,Tieri:preprint,Debnath:2018}: In a homogeneous medium, SSR consists in the buildup of a stable macroscopic dipole, that acts as a stationary source of coherent light. The dynamical properties can be understood in terms of a peculiar time crystal \cite{Tucker:2018}, which locks at a frequency determined by the incoherent pump rate $w$. In a homogeneous medium the transition from normal to SSR fluorescence is controlled by $w$ when the superradiant decay rate is larger than the rates characterizing other incoherent processes.  Here, we show that in the presence of the optomechanical coupling with the external degrees of freedom SSR corresponds to spatio-temporal long-range order and is reached when the characteristic rates exceed the recoil frequency, scaling the mechanical energy exchanged with radiation. When instead the recoil frequency becomes comparable with the pump or the superradiant decay rate, then the superradiant emitted light can become either chaotic or incoherent. The chaotic phase, in particular, characterizes the asymptotic phase of an incoherent dynamics, it emerges from the interplay between quantum fluctuations, noise, and global interactions mediated by the cavity field, and is thus qualitatively different from chaos reported in quantum dynamics of Hamiltonian global-range interacting systems \cite{Emary:2003,Lerose:2018}.

Consider a gas of $N$ atomic bosons with mass~$m$ that are confined along the axis of a standing-wave resonator. The atoms do not interact directly; their relevant electronic degrees of freedom form a metastable dipole with excited state~$\ket{e}$ and ground state~$\ket{g}$. The dipoles are incoherently pumped at rate~$w$ and strongly coupled to a cavity mode with wave number~$k$ and loss rate~$\kappa$.  The evolution of the density matrix~$\hat{\varrho}$ for the cavity field and the atoms' internal and external degrees of freedom is given by the Born-Markov master equation $\partial_t\hat{\varrho}=[\hat{H}_0+\hat{H}_c,\hat{\varrho}]/(i\hbar)+w\sum_j\mathcal{L}[\hat{\sigma}_j^\dagger]\hat{\varrho}+\kappa\mathcal{L}[\hat{a}]\hat{\varrho}$. Here, $\hat{H}_0=\sum_{j=1}^N\hat{p}_j^2/(2m)$ is the total kinetic energy, with $\hat{p}_j$ the momentum of each atom $j$; $\hat{H}_c=\hbar\Delta \hat{a}^\dagger \hat{a} +\hbar gN(\hat{a}^\dagger \hat{X}/2+{\rm H.c.})$ describes the reversible evolution due to the interaction with the resonator, with $\hat{a}$ and $\hat{a}^\dagger$ the annihilation and creation operators of a cavity photon, and~$\Delta$ the cavity detuning from the atomic transition frequency. The field couples with strength~$g$ to the collective dipole $\hat{X}=\sum_j\hat{\sigma}_j\cos(k\hat{x}_j)/N$, where $\hat{\sigma}_j=\ket{g}_j\bra{e}$ 
and the sum is weighted by the value of the cavity standing-wave mode $\cos(kx)$ at the positions $\hat x_j$. The Lindbladians describe the incoherent dynamics and read $
\mathcal{L}[\hat{O}]\hat{\varrho}=-\left(\hat{O}^{\dag}\hat{O}\hat{\varrho}+\hat{\varrho}\hat{O}^{\dag}\hat{O}\right)/2+\hat{O}\hat{\varrho}\hat{O}^{\dag}$.
For $N\gg1$ the quantum dynamics is numerically intractable due to the adverse Liouville space scaling. This dynamics can be cast in terms of long-range dipolar and optomechanical interactions in the  atoms' Hilbert space when~$\kappa$ and $\Delta$ are the largest rates. In this regime the atomic transition is radiatively broadened by the coupling with the cavity, its linewidth at an antinode is $\Gamma_c=g^2\kappa/(\kappa^2+4\Delta^2)$. Then, the cavity field follows adiabatically the atomic motion, $\hat{a}\propto \hat{X}$ \cite{Xu:2016,Jaeger:2017}, while shot-noise fluctuations are negligible  \cite{Habibian:2013}. The atoms density matrix~$\hat{\rho}_N$ then obeys the master equation $\partial_t\hat{\rho}_N=[\hat{H}_{\rm eff},\hat{\rho}_N]/(i\hbar)+w\sum_j\mathcal{L}[\hat{\sigma}_j^\dagger]\hat{\rho}_N+\Lambda\mathcal{L}[\hat{X}]\hat{\rho}_N$. Here, $\hat{H}_{\mathrm{eff}}=\hat{H}_0+\hat{V}$,
where $\hat{V}=-\hbar N\Lambda(\Delta/\kappa)\hat{X}^{\dag}\hat{X}$ describes the global interactions mediated by cavity photons. Now the incoherent processes are the incoherent pump at rate~$w$ and the superradiant decay with rate~$\Lambda=N\Gamma_c$. We neglected retardation effects of the cavity field, which is justified by the choice of large~$\kappa$. We also neglected single-atom radiative decay at rate~$\Gamma_c$, assuming time scales $t<1/\Gamma_c$ and $N\gg1$. Since~$1/\Gamma_c=N/\Lambda$, this time scale can be stretched to $t\to\infty$ in a thermodynamic limit $N\to\infty$ where $\Lambda$ is kept constant \cite{Emary:2003,Jaeger:2017}.  Under these assumptions we finally obtain the mean-field master equation for the single-particle density matrix $\hat{\rho}_1$ (assuming that $\hat\rho_N$ is a product state at $t=0$):
\begin{equation}
\label{MF:MEq}
\partial_t\hat{\rho}_{1}=[\hat{H}_{\mathrm{mf}}\{\hat{\rho}_1\},\hat{\rho}_{1}]/(i\hbar)+w\mathcal{L}[\hat{\sigma}^{\dag}]\hat {\rho}_{1}\,,
\end{equation} 
where $\hat{\rho}_1={\rm Tr}_{N-1}\{\hat\rho_N\}$ is obtained by tracing out ${N-1}$~atoms.
Now the incoherent evolution is due entirely to the incoherent pump and the interactions with the resonator are given by the mean-field Hamiltonian:
\begin{align}
\hat{H}_{\mathrm{mf}}=&\frac{\hat{p}^2}{2m}-\frac{\hbar\Lambda}{2\sin\chi}\left(e^{i\chi} X\{\hat{\rho}_1\}\hat{\sigma}^{\dag}+{\rm H.c.}\right)\cos(k\hat{x})\,, \label{H2}
\end{align}
with $\tan(\chi)=\kappa/(2\Delta)$. Here, the Rabi frequency is proportional to the mean-field order parameter $X\{\hat{\rho}_1\}=\mathrm{Tr}\{\hat{\sigma}\cos(k\hat{x})\hat{\rho}_1\}$, and thus depends on the global macroscopic dipole. Note that $X$ generates the intracavity field and within the mean-field treatment determines the field's coherence properties. By neglecting the diffusion due to the incoherent pump, Eq.~\eqref{MF:MEq} can be reduced to a Vlasov equation with a potential that depends on the macroscopic dipole of the initial state, and whose stable solutions are metastable states of the out-of-equilibrium dynamics \cite{Campa:2009,Levin:2014}. In the following we analyze the stability of a thermal initial state $\hat{\rho}_1^{\rm (0)}=|e\rangle\langle e| \otimes \exp (-\beta \hat{p}^2/2m)/Z$, with inverse temperature $\beta$, partition function $Z$. Here, $X\{\hat{\rho}_1^{\rm (0)}\}=0$.
\begin{figure}[h!]
	\center \includegraphics[width=0.8\linewidth]{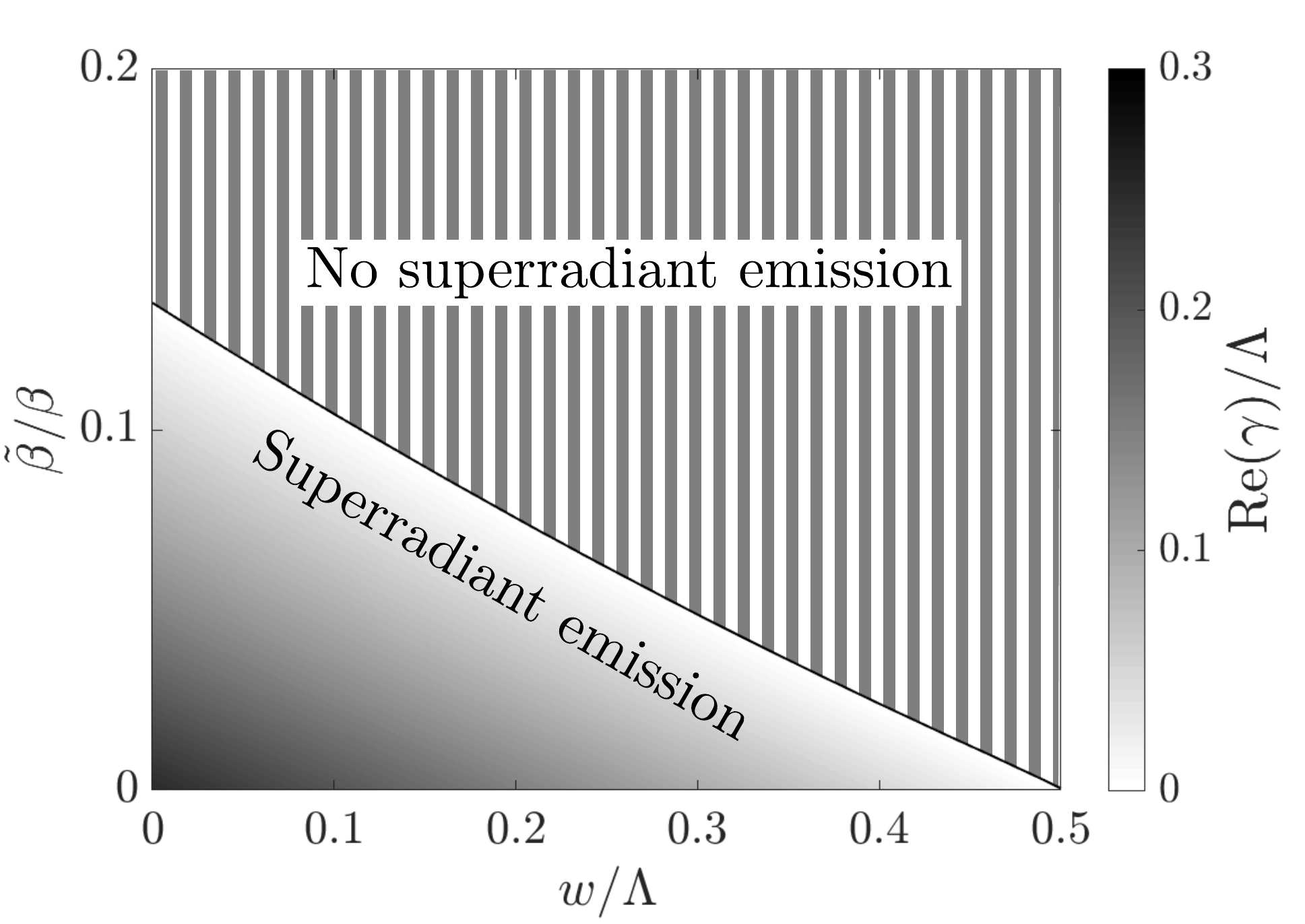}
	\caption{Contour plot of the rate $\gamma$ of the first superradiant emission as a function of the incoherent pump rate $w$ (in units of~$\Lambda$) and of the atomic gas temperature $1/\beta$ (in units of $\tilde{\beta}^{-1}=\hbar \Lambda^2/(2\omega_R)$). The solid line separates the regime in which the atoms undergo superradiant decay from the one where thermal fluctuations suppress superradiance (stripes). \label{Fig:2}}
\end{figure}

The short-time dynamics is determined by means of a stability analysis as a function of $w$ and $\beta$, see Supplemental Material~(SM)~\cite{SM} for details. No superradiant emission is found when $X\{\hat{\rho}_1^{\rm (0)}\}=0$ is stable to small fluctuations. When instead exponentially increases as $X\sim\exp(\gamma t)$ with $\mathrm{Re}(\gamma)>0$, then the system undergoes superradiant decay with $\mathrm{Re}(\gamma)$. Figure~\ref{Fig:2} shows the contour plot of the exponent~$\mathrm{Re}(\gamma)$ as a function of both $w$ and $\beta$. We find a threshold temperature $k_BT_c\approx 0.1\hbar \Lambda^2/(2\omega_R)$, where~$\omega_R=\hbar k^2/(2m)$ is the recoil frequency. For $T>T_c$ thermal fluctuations suppress superradiance. For $T<T_c$ superradiance is found for a finite interval of the pump rate $0< w\le w_{\rm max}(\beta)$, which increases with the ratio $\eta=\beta/\bar{\beta}=T_c/T$. For $\eta\to\infty$ the upper bound is $w_{\rm max}=\Lambda/2$, that coincides with the value found for a homogeneous medium \cite{Jaeger:2017}. We now focus on the regime where $\Lambda$ is of the order of $\omega_R$, so that the threshold temperature $T_c$ can be several $\mu K$.
\begin{figure}[h!]
	\center \includegraphics[width=0.8\linewidth]{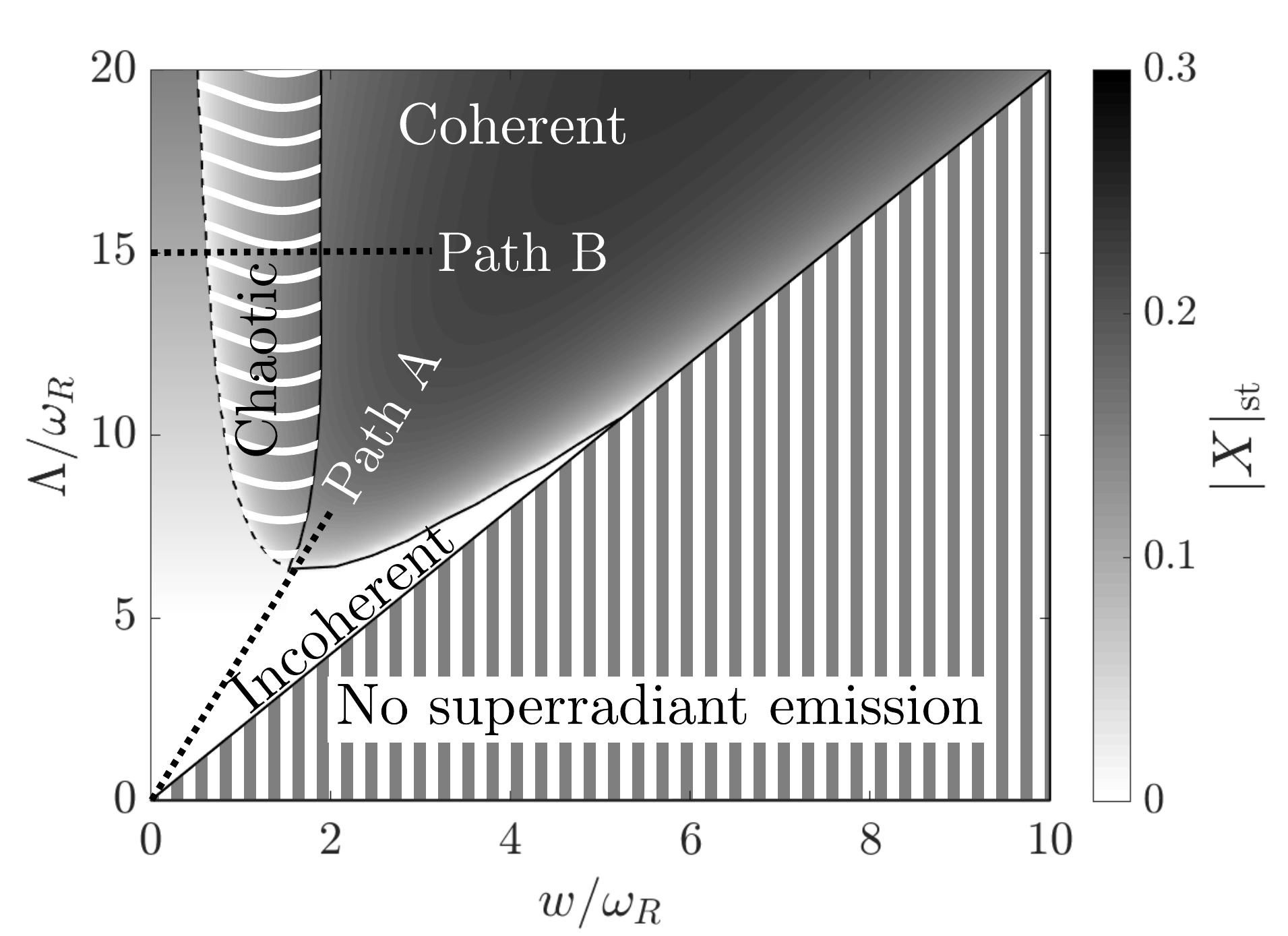}
	\caption{Phase diagram in the $w/\omega_R$--$\Lambda/\omega_R$ plane when the atoms initially form a Bose-Einstein condensate at $T=0$. The phases are labeled by the coherence properties of the emitted light. The emitted field is given by $X(t)$ and is obtained by solving Eq.~\eqref{MF:MEq} at the asymptotic dynamics, see \cite{SM}. Path A (Path B) shows the parameters of Fig.~\ref{Fig:4} (Fig.~\ref{Fig:5}). In the striped region superradiant decay is suppressed (corresponding to the region at $T=0$ and $w>\Lambda/2$ in Fig.~\ref{Fig:2}).  \label{Fig:3}}
\end{figure}

We now study the dynamics of an ensemble of atoms in the zero-temperature limit, when the atoms initially form a Bose-Einstein condensate (BEC). We neglect onsite interactions and analyze the dynamics of the external degrees of freedom on the closed family of momentum states $|\Psi_0\rangle=|0\rangle$ (the BEC) and $|\Psi_n\rangle=(|n \hbar k\rangle+|-n \hbar k\rangle)/\sqrt{2}$ $(n=1,2,\ldots)$. These states are coupled by absorption and emission of cavity photons; their energy $E_{{\rm kin},n}=n^2\hbar\omega_R$ is an integer multiple of $\omega_R$. The asymptotic behavior of Eq.~\eqref{MF:MEq} is strictly defined in the thermodynamic limit and is determined by means of a recursive procedure \cite{SM}. In Fig.~\ref{Fig:3} we report the coherence properties of the emitted light in a $w-\Lambda$ phase diagram. We first note the normal (striped) phase with $w>\Lambda/2$, where there is no superradiant emission. The transition from normal to superradiant phase (without optomechanical coupling) has been discussed in the literature \cite{Meiser:2009,Meiser:2010:1,Meiser:2010:2,Tucker:2018,Barberena:2018}. Within the regime where SSR is expected, we now find that the optomechanical coupling  gives rise to three phases which we denote by (i) incoherent, (ii) coherent, and (iii) chaotic, corresponding to the coherence properties of the emitted light. In the incoherent phase only the solution with $X=0$ is stable and collective effects are suppressed. In the coherent phase there is one stable solution with $X\neq 0$. As visible in the phase diagram, the condition for the appearance of this phase is that the superradiant linewidth exceeds a minimum value determined by the recoil frequency, $\Lambda>\Lambda_c$ with $\Lambda_c\sim 6\omega_R$. Finally, the chaotic phase is found for $\Lambda>\Lambda_c$, when the pump rate is below a threshold $w_c(\Lambda)$. Here, both solutions with $X\neq0 $ and $X=0$ are unstable. 

\begin{figure}[h!]
	\center \includegraphics[width=0.9\linewidth]{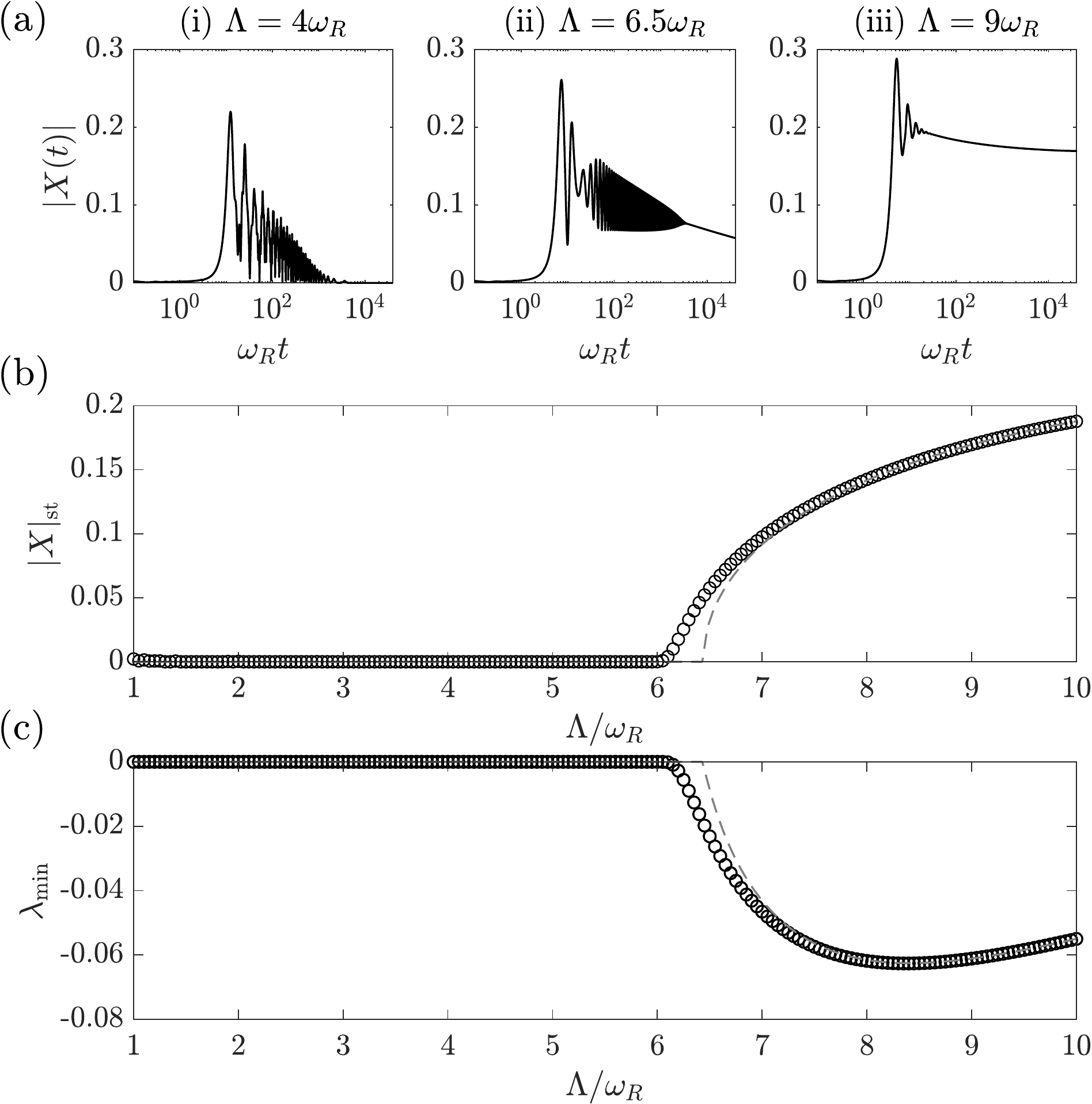}
	\caption{The incoherent-coherent transition for the parameters of Path A of Fig.~\ref{Fig:3} ($w=\Lambda/4$ and $\Delta=\kappa/2$). Subplot (a), from left to right: Dynamics of $X$ for $\Lambda=4,6.5,9\omega_R$. (b) The asymptotic value for the mean-field order parameter $|X(t_{\mathrm{f}})|$ and (c) the minimum eigenvalue $\lambda_{\rm min}$ of the partial transpose of the asymptotic density matrix, signalling entanglement between external and internal degrees of freedom, as a function of $\Lambda$ (in units of $\omega_R$). Black circles: Numerical results at time $t_{\mathrm{f}}=4\times 10^4\omega_R^{-1}$; Dashed lines: Steady-state values from the iterative solution of $\partial_t \rho_1=0$, Eq.~\eqref{MF:MEq}. 
\label{Fig:4}}
\end{figure}

We verified these predictions by numerically integrating Eq.~\eqref{MF:MEq} with the initial state $\rho_1{(0)}$ at $T=0$ on the grid of momentum states $p=0,\pm \hbar k,  \ldots,\pm 15\hbar k$. Figure~\ref{Fig:4}(a) displays 
$|X(t)|$ for different values of $\Lambda$ along Path A of Fig.~\ref{Fig:3}, where a direct transition occurs from an incoherent to a coherent (SSR) phase. For all values the intracavity field $|X(t)|$ first grows exponentially, and subsequently reaches a maximum at a time scale $\tau_c\sim 1/\Lambda$. After this time scale: (i) For $\Lambda<\Lambda_c$ the intracavity field $|X(t)|$ decays to zero. This dynamics is accompanied by the formation of a statistical mixture of states $\ket{e,\Psi_{2n}}$ and  $|e,\Psi_{2n+1}\rangle$, which dephases the macroscopic dipole and leads to suppression of superradiant emission. (ii) For $\Lambda \sim \Lambda_c$ the field undergoes fast oscillations and then slowly decays to zero. (iii) For $\Lambda>\Lambda_c$ the field oscillates about a finite asymptotic value and the atoms form a stable spatial pattern. This dynamics exhibits the general features of a dynamical phase transition, which occurs after the first superradiant emission at $t\sim \tau_c$. After $\tau_c$ the macroscopic dipole $X$ decays to zero or oscillates about a finite metastable value. We denote the asymptotic value of the order parameter by $X_{\rm st}(\Lambda)$, which we determine by numerical evolution of $|X(t)|$, taking $|X_{\rm st}(\Lambda)|=|X(t_{\mathrm{f}})|$, where at $t_\mathrm{f}$ the dipole $|X(t)|$ has reached a constant value. We compare this result with the asymptotic solution $\hat\rho_{\rm st}$ of Eq.~\eqref{MF:MEq}, using an iterative procedure based on a seed $X>0$ (as for determining the phase diagram of Fig.~\ref{Fig:3}~\cite{SM}). Along Path A this iterative procedure always converges to either $X_{\rm st}=0$ for $\Lambda<\Lambda_c$ and $X_{\rm st}>0$ for $\Lambda>\Lambda_c$. As is visible in Fig.~\ref{Fig:4}(b), the predictions obtained by numerical integration (circles) and by the iterative procedure (dashed line) qualitatively agree and exhibit the features of a second-order phase transition. Figure~\ref{Fig:4}(c) displays the minimum eigenvalue of the partial transpose of $\hat\rho_{\rm st}$. Its behavior shows that at the buildup of SSR internal and external degrees of freedom become entangled \cite{SM}.  

\begin{figure}[h!]
	\center \includegraphics[width=0.8\linewidth]{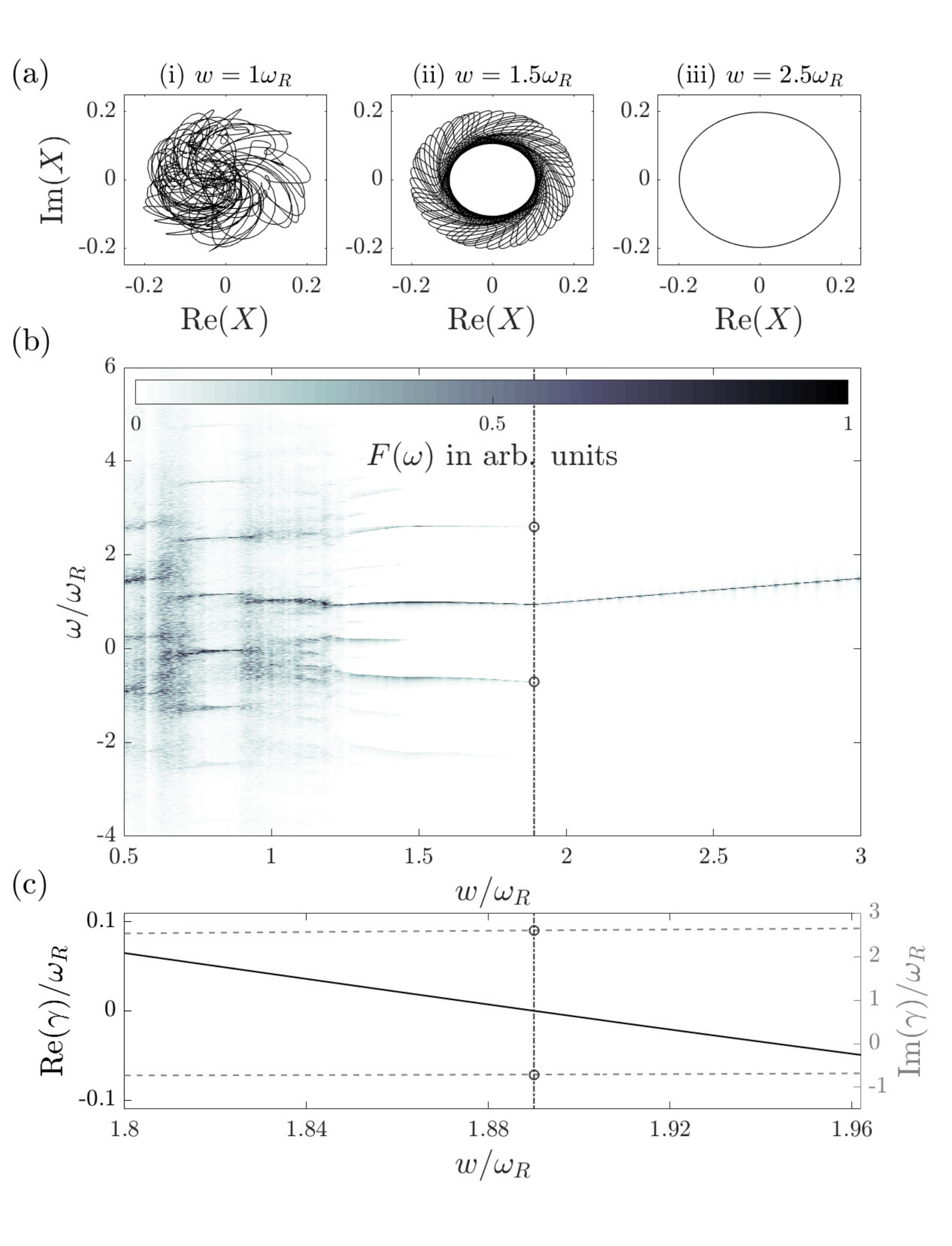}
	\caption{The chaotic-coherent transition for the parameters of Path B of Fig.~\ref{Fig:3} ($\Lambda=15\omega_R$ and $\Delta=\kappa/2$). (a) From left to right: Real and imaginary part of $X$ for $w=1,1.5,2.5\omega_R$ (here for the time interval $t\in[9.8\times10^3,10^4]/\omega_R$). (b) Contour plot of the spectrum of the emitted light $F(\omega)$ (arbitrary units) as a function of $w$ and of the frequency $\omega$ (in units of $\omega_R$). Here,  $F(\omega)\propto\left|\int_0^{t_{\mathrm{end}}}e^{i\omega t}X(t)dt\right|$ is found by integrating Eq.~\eqref{MF:MEq} until $t_{\mathrm{end}}=10^4 \omega_R^{-1}$. 
	(c) The real (solid) and imaginary part (dashed) of the exponent $\gamma$ (in units of $\omega_R$) giving the stability of the stationary solutions. The vertical dashed lines indicate the critical pumping strength $w_c(\Lambda)$, where $\mathrm{Re}(\gamma)$ changes sign and the sidebands appear, the circles mark the corresponding frequencies. \label{Fig:5} }
\end{figure}

The transition separating the coherent from the chaotic phase occurs for $\Lambda>\Lambda_c$ as a function of $w$: The properties of the emitted light dramatically depend on whether $w$ is smaller or larger than a critical value $w_c(\Lambda)$. Figure~\ref{Fig:5}(a) displays the numerical results for the real and the imaginary part of $X(t)$ for a fixed time interval for (i) $w<w_c$, where the dynamics is chaotic, (ii) $w\simeq w_c$ where the dynamics is mainly characterized by the appearance of two subharmonics, and (iii) for $w>w_c$, where the dynamics is evidently coherent. The spectrum of the emitted light is displayed in Fig.~\ref{Fig:5}(b) as a function of $w$ and for the parameters of Path B of Fig.~\ref{Fig:3}. The transition from regular oscillations to chaos occurs at a value $w_c$ where two sidebands appear. We analytically determine $w_c$ by means of a stability analysis, see \cite{SM}. This analysis also delivers the frequencies of the sideband at $w=w_c$ and the Lyapunov exponent $\gamma_L={\rm Re}(\gamma)$. As is visible in Fig.~\ref{Fig:5}(c), $\gamma_L$ changes sign at $w=w_c$ and is positive for $w<w_c$. The trajectory of subplot (a)-(i) corresponds to the value of $w$ where the spectrum is dense: In this parameter regime the stability analysis predicts the transition from chaotic to incoherent dynamics. Numerical simulations show that for $w<w_c$ the density grating becomes unstable and the system jumps back and forth between a prevailing occupation of the set of states corresponding to an even grating, $\{\ket{e,\Psi_{2n}},\ket{g,\Psi_{2n+1}},n=0,1,2,...\}$, and of the ones corresponding to an odd grating, $\{\ket{e,\Psi_{2n+1}},\ket{g,\Psi_{2n}},n=0,1,2,...\}$. While the states within each set are coupled by coherent processes, the two sets are only coupled to each other by the incoherent pump: Thus, for $w<w_c$ the long-range optomechanical interactions tend to form a grating, which locks the phase of the field, while the incoherent pump induces quantum jumps between different gratings.
An analysis of the entanglement is possible only from the coherent side, where the non-linear master equation has one stationary solution, and shows that internal and external degrees of freedom are entangled for $w>w_c$.  We remark that in the coherent phase the frequency of the oscillator depends on the incoherent pump rate, $\omega_a+\Delta w/\kappa$, showing that this spatio-temporal selforganization exhibits the features of time crystals \cite{Tucker:2018}. 
	
The phase diagram can be observed by tuning the superradiant linewidth and the pump rate across values of the order of the recoil frequency $\omega_R$, the phases are signaled by the first-order correlation function of the emitted light. These dynamics can be realized when the resonator linewidth $\kappa$ exceeds by several orders of magnitude $\omega_R$ and when other incoherent processes can be discarded over the time scales where the dynamical phase transition occurs. Specifically, the spontaneous decay of the dipolar transition shall be orders of magnitude smaller than the recoil frequency, which can be realized using a Raman transition between metastable hyperfine states, as for instance in Refs.~\cite{Bohnet:2012,Kroeze:2018,Landini:2018}. 
\acknowledgements
The authors are grateful to J.~Eschner, L.~Giannelli, F.~Rosati,  S.~Sch\"utz, and M.~Xu for discussions. This work has been supported by the German Research Foundation (DACH ``Quantum crystals of matter and light'' and the priority programme no. 1929 GiRyd), by the European Commission (ITN "ColOpt") and by the German Ministry of Education and Research (BMBF) via the QuantERA project ``NAQUAS''. Project NAQUAS has received funding from the QuantERA ERA-NET Cofund in Quantum Technologies implemented within the European Union's Horizon 2020 programme. 

\textit{Note added.} After submission of this work we became aware of Refs.~\cite{Dogra:2019,Chiacchio:2019} where analogous dynamics in similar setups are studied.

\newpage
\setcounter{equation}{0}
\setcounter{figure}{0}

\renewcommand*{\citenumfont}[1]{S#1}
\renewcommand*{\bibnumfmt}[1]{[S#1]}
\renewcommand{\thesection}{S.\arabic{section}}
\renewcommand{\thesubsection}{\thesection.\arabic{subsection}}
\makeatletter 
\def\tagform@#1{\maketag@@@{(S\ignorespaces#1\unskip\@@italiccorr)}}
\makeatother
\makeatletter
\makeatletter \renewcommand{\fnum@figure}
{\figurename~S\thefigure}
\makeatother
\onecolumngrid
\begin{center}
	\textbf{\large Supplemental Material: Dynamical phase transitions to optomechanical superradiance}
\end{center}
\twocolumngrid

\subsection{Stability analysis}
In this section, we study the stability of a stationary state of the mean-field master equation 
\begin{align}
	\frac{\partial\hat{\rho}_1}{\partial t}=\mathcal{L}_{\mathrm{mf}}[\hat{\rho}_1]\hat{\rho}_1.\label{mastereq}
\end{align}
The explicit form of $\mathcal{L}_{\mathrm{mf}}[\hat{\rho}_1]$ is given in Eq.~(1) of the main article. The stability of a stationary state $\hat{\rho}_0$, with $\partial_t\hat{\rho}_0=0$, is determined by the initial dynamics of a density matrix $\hat{\rho}=\hat{\rho}_0+\delta\hat{\rho}$ with a small perturbation $\delta\hat{\rho}$ \cite{Campa:2009}. If this perturbation is amplified over time we can state that $\hat{\rho}_0$ is unstable, otherwise $\hat{\rho}_0$ is stable. 

Using the mean-field master equation \eqref{mastereq} we derive an equation of motion for $\delta\hat{\rho}$ that takes the form
\begin{align}
	\frac{\partial \delta\hat{\rho}}{\partial t}=i\frac{\Lambda}{2}\left(\alpha \delta X^*[\hat{J}_1,\hat{\rho}_0]-\mathrm{H.c.}\right)+\mathcal{L}_{\mathrm{mf}}[\hat{\rho}_0]\delta\hat{\rho}.\label{Stab:1}
\end{align}
Here we have defined $\alpha=2\Delta/\kappa-i$ and used the definition
\begin{align}
	\mathcal{L}_{\mathrm{mf}}[\hat{\rho}_0]\delta\hat{\rho}=\frac{1}{i\hbar}[\hat{H}_{\mathrm{mf}}[\hat{\rho}_0],\delta\hat{\rho}_1]+w\mathcal{L}[\hat{\sigma}^{\dag}]\delta\hat{\rho},\label{Lmf}
\end{align}
with $\hat{J}_1=\hat{\sigma}\cos(k\hat{x})$ and $\delta X=X\{\delta\hat{\rho}\}$. In Eq.~\eqref{Stab:1}, we have included only first order perturbations in $\delta\hat{\rho}$ and we have discarded the second order. 
Applying the Laplace transform $\mathrm{L}[f](s)=\int_{0}^\infty\,dt\,e^{-st}f(t)$ we derive the following equation 
\begin{align*}
	{\bf D}(s)\begin{pmatrix}
		\mathrm{L}[\delta X](s)\\
		\mathrm{L}[\delta X^*](s)
	\end{pmatrix}=\begin{pmatrix}
		\mathrm{Tr}\left(\hat{J}_1[s-\mathcal{L}_{\mathrm{mf}}[\hat{\rho}_0]]^{-1}\delta\hat{\rho}(0)\right)\\
		\mathrm{Tr}\left(\hat{J}_1^{\dag}[s-\mathcal{L}_{\mathrm{mf}}[\hat{\rho}_0]]^{-1}\delta\hat{\rho}(0)\right)
	\end{pmatrix}
\end{align*}
with
\begin{align}
	{\bf D}(s)=&\begin{pmatrix}
		1+C_{11}(s)&C_{12}(s)\\
		C_{21}(s)&1+C_{22}(s)\\
	\end{pmatrix}.\label{Cmatrix}
\end{align}
The entries of the matrix take the forms
\begin{align}
	C_{11}=&-i\frac{\Lambda}{2}\alpha^*\mathrm{Tr}\left(\hat{J}_1\left(s-\mathcal{L}_{\mathrm{mf}}[\hat{\rho}_0]\right)^{-1}[\hat{J}_1^{\dag},\hat{\rho}_0]\right),\\
	C_{12}=&-i\frac{\Lambda}{2}\alpha\mathrm{Tr}\left(\hat{J}_1\left(s-\mathcal{L}_{\mathrm{mf}}[\hat{\rho}_0]\right)^{-1}[\hat{J}_1,\hat{\rho}_0]\right),\\
	C_{21}=&-i\frac{\Lambda}{2}\alpha^*\mathrm{Tr}\left(\hat{J}_1^{\dag}\left(s-\mathcal{L}_{\mathrm{mf}}[\hat{\rho}_0]\right)^{-1}[\hat{J}_1^{\dag},\hat{\rho}_0]\right),\\
	C_{22}=&-i\frac{\Lambda}{2}\alpha\mathrm{Tr}\left(\hat{J}_1^{\dag}\left(s-\mathcal{L}_{\mathrm{mf}}[\hat{\rho}_0]\right)^{-1}[\hat{J}_1,\hat{\rho}_0]\right).
\end{align}
Inverting ${\bf D}(s)$ and applying the inverse Laplace transformation, we obtain the dynamics of $\delta X$. To calculate the dynamics we need to know the poles when we invert the matrix~${\bf D}(s)$. These are roots of the dispersion relation
\begin{align}
	D(s)=\det({\bf D}(s))=0.\label{Dispersion}
\end{align} 
The complex solution $\gamma$ of Eq.~\eqref{Dispersion} 
with the largest real part $\mathrm{Re}(\gamma)$ gives the dominant contribution to the dynamics of $\delta X$. Therefore this determines whether the stationary solution $\hat{\rho}_0$ is stable or not. If $\mathrm{Re}(\gamma)>0$ the perturbation $\delta\hat{\rho}$ will exponentially grow and thus $\hat{\rho}_0$ is an unstable stationary solution. Otherwise, if $\mathrm{Re}(\gamma)\leq0$, $\hat{\rho}_0$ is stable.

\subsection{Asymptotic state}
In this section we explain how we calculate the stationary state of the system leading to the diagram in Fig.~3 of the main article.

A significant class of stationary states are given by incoherent states 
\begin{align}
	\hat{\rho}_0=\hat{\rho}_{\mathrm{mom}}\otimes|e\rangle\langle e|.\label{incoherent}
\end{align} 
These are referred to as incoherent since the collective dipole $X\{\hat{\rho}_0\}=0$ vanishes. These states are stationary if they commute with the kinetic energy $[\hat{\rho}_{\mathrm{mom}},\hat{p}^2]=0$. Although these states do not show a collective dipole they can be used to calculate the onset of superradiance. This is presented in section ``{\bf Stability of the incoherent state}''. In this section we explain how we find stationary states that show a non-vanishing collective dipole $X\{\hat{\rho}_0\}\neq0$.

We will show that there is a stationary state where $|X|\neq0$ and $\langle\hat{p}^2\rangle$ is not time dependent while $X=|X|e^{i\phi(t)}$ oscillates with a constant frequency in time.

Using Eq.~\eqref{mastereq} one can show that 
\begin{align}
	\frac{d}{dt}\langle\hat{H}_{\mathrm{mf}}\rangle=&\frac{d}{dt}\left(\frac{\langle \hat{p}^2\rangle}{2m}-\frac{\hbar\Lambda\Delta}{\kappa/2}|X|^2\right)\nonumber\\
	=&\mathrm{Tr}\left(\hat{H}_{\mathrm{mf}}\frac{\partial\hat{\rho}_1}{\partial t}\right)+\mathrm{Tr}\left(\frac{\partial \hat{H}_{\mathrm{mf}}}{\partial t}\hat{\rho}_1\right)\nonumber\\
	=&\frac{\hbar \Lambda\Delta}{\kappa}w|X|^2-\frac{\hbar \Lambda }{2}\left(\alpha^*\frac{dX}{dt}X^*+\mathrm{c.c}\right).\label{calculatingphi}
\end{align}
Explicitly denoting the amplitude and phase $X=|X|e^{i\phi}$ we obtain
\begin{align}
	\frac{d}{dt}\langle\hat{H}_{\mathrm{mf}}\rangle=\frac{\hbar \Lambda\Delta }{\kappa}w|X|^2+\hbar\Lambda|X|^2\frac{d\phi}{dt}-\frac{\hbar\Lambda\Delta}{\kappa}\frac{d|X|^2}{dt}.\label{calculatingphi2}
\end{align}
Now assuming that there exists a stationary state with $d\langle \hat{p}^2\rangle/dt=0$ and $d|X|^2/dt=0$ we arrive at
\begin{align}
	\frac{d\phi}{dt}=-\frac{w\Delta }{\kappa}.\label{omega0}
\end{align}
Therefore to find a stationary solution for the system we need to solve
\begin{align}
	\mathcal{L}_{\mathrm{mf}}\left[\hat{\rho}_1\right]\hat{\rho}_1=\frac{1}{i\hbar}\left[\frac{\hbar w\Delta }{\kappa}\hat{\sigma}^{\dag}\hat{\sigma},\hat{\rho}_1\right].\label{fixpoint}
\end{align}
This is equivalent to calculating the stationary state in the frame oscillating with the frequency shown in Eq.~\eqref{omega0}.

To characterize and numerically determine this solution we use the order parameter $X$. For the numerical calculation of the stationary state we start from an order parameter $X>0$ and find $\hat{\rho}_0$ to recalculate the new value of $X=\mathrm{Tr}(\hat{\sigma}\cos(k\hat{x})\hat{\rho}_0)$. We iterate this step until $\hat{\rho}_0$ and $X$ converge. 

In the case when there is a solution of Eq.~\eqref{fixpoint} with $X\neq0$ we know that there is a coherent stationary state. However, this state does not need to be stable. To calculate the stability we use the dispersion relation in Eq.~\eqref{Dispersion}. 
\subsection{Stability of the incoherent state}
The aim of this section is to describe the stability of the incoherent state in Eq.~\eqref{incoherent}. This method leads to the stability diagram of a thermal state visible in Fig.~2 of the main article. 

Since $X\{\hat{\rho}_0\}=0$ we observe that the matrix in Eq.~\eqref{Cmatrix} becomes diagonal. Therefore if we want to find the zeros of the dispersion relation in Eq.~\eqref{Dispersion} it is sufficient to solve the equation
\begin{align}
	1+C_{11}(s)=0.\label{1+c11}
\end{align}
Using the duality of the Schr\"odinger and Heisenberg pictures we obtain
\begin{align}
	C_{11}=&-i\frac{\Lambda}{2}\alpha^*\mathrm{Tr}\left(\hat{J}_1\left(s-\mathcal{L}_{\mathrm{mf}}\right)^{-1}[\hat{J}_1^{\dag},\hat{\rho}_0]\right)\nonumber\\
	=&-i\frac{\Lambda}{2}\alpha^*\int_{0}^{\infty} dte^{-st}\mathrm{Tr}\left(\hat{J}_1e^{\mathcal{L}_{\mathrm{mf}}t}[\hat{J}_1^{\dag},\hat{\rho}_0]\right)\nonumber\\
	=&-i\frac{\Lambda}{2}\alpha^*\int_{0}^{\infty} dte^{-st}\langle[\hat{J}_1(t),\hat{J}_1^{\dag}(0)]\rangle_{\hat{\rho}_0}.\label{susz}
\end{align}
Here we use the definition that for an operator $\hat{A}$ the expectation value is defined as $\langle \hat{A}\rangle_{\hat{\rho}}=\mathrm{Tr}(\hat{A}\hat{\rho})$.

In the homogeneous case, we calculate $\hat{J}_1(t)$, and it takes the form
\begin{align}
	\hat{J}_1(t)=\hat{\sigma}(0)e^{-\frac{w}{2}t}\cos\left(k\hat{x}(t)\right), \label{J1hom}
\end{align}
with $\hat{x}(t)=\hat{x}(0)+k\hat{p}(0)t/m$.
Using Eq.~\eqref{J1hom} in Eq.~\eqref{susz} we obtain
\begin{align}
	C_{11}=i\frac{\Lambda}{2}\alpha^*\int_{0}^{\infty} dte^{-\left(s+\frac{w}{2}\right)t}\left\langle\cos(k\hat{x})\cos\left(k\hat{x}(t)\right)\right\rangle_{\hat{\rho}_{\mathrm{mom}}},
\end{align}
where we explicitly used the fact that all particles are in the excited state and therefore $\langle\hat{\sigma}\hat{\sigma}^{\dag}\rangle=0$ holds. From the identity
\begin{align*}
	e^{ik\hat{x}+ik\hat{p}/mt}=e^{ik\hat{x}}e^{ik\hat{p}/mt}e^{i\omega_Rt}
\end{align*}
and momentum translation
\begin{align*}
	e^{ik\hat{x}}|p\rangle=|p+\hbar k\rangle
\end{align*}
we can show that
\begin{align}
	I&=\int dp\bigg\langle p\bigg|\cos(k\hat{x})\frac{e^{ik\hat{x}}e^{ik\hat{p}/mt}+e^{-ik\hat{x}}e^{-ik\hat{p}/mt}}{2}\hat{\rho}_{\mathrm{mom}}\bigg|p\bigg\rangle\nonumber\\
	&=\frac{1}{4}\int dp\langle p|\hat{\rho}_{\mathrm{mom}}|p\rangle (e^{ikp/mt}+e^{-ikp/mt})\nonumber\\
	&+\frac{1}{4}\left(\langle \hbar k\big|\hat{\rho}_{\mathrm{mom}}|-\hbar k\rangle +\langle -\hbar k\big|\hat{\rho}_{\mathrm{mom}}|\hbar k\rangle\right)e^{-i2\omega_Rt}.\label{I}
\end{align}
Here, it is necessary that the condition $\langle p|\hat{\rho}_{\mathrm{mom}}|p'\rangle\neq0$ can only hold for $p'=\pm p$. This is true since $\hat{\rho}_{\mathrm{mom}}$ needs to commute with $\hat{p}^2$ that $\hat{\rho}_0=\hat{\rho}_{\mathrm{mom}}\otimes|e\rangle\langle e|$ is a stationary state. Using
\begin{align}
	C_{11}=i\frac{\Lambda}{2}\alpha^*\int_{0}^{\infty} dte^{-\left(s+\frac{w}{2}-i\omega_R\right)t}I
\end{align}
and Eq.~\eqref{I} we get 
\begin{align}
	C_{11}=&i\frac{\Lambda y}{4}\alpha^*\int_{-\infty}^{\infty} dp\frac{\langle p|\hat{\rho}_{\mathrm{mom}}|p\rangle}{y^2+\left(\frac{kp}{m}\right)^2}\nonumber\\
	&+i\frac{\Lambda}{8}\alpha^*\frac{\langle \hbar k\big|\hat{\rho}_{\mathrm{mom}}|-\hbar k\rangle +\langle -\hbar k\big|\hat{\rho}_{\mathrm{mom}}|\hbar k\rangle}{y+i2\omega_R},\label{C11finalhom}
\end{align}
with $y=s+w/2-i\omega_R$. 

Figure~2 of the main article shows $\mathrm{Re}(\gamma)$ for a thermal state $\hat{\rho}_{\mathrm{mom}}=\exp (-\beta \hat{p}^2/2m)/Z$ as function of $\beta$ and $w/\Lambda$. The value of $\gamma$ is found by solving numerically Eq.~\eqref{1+c11} using Eq. \eqref{C11finalhom}. 
\subsection{Entanglement}
In this subsection we report how we calculate the smallest eigenvalue $\lambda_{\mathrm{min}}$ of the partial transpose of the density matrix $\hat{\rho}$ that is shown for the transition from incoherent to coherent in Fig.~4(c). 
\begin{figure}[h!]
	\flushleft(a)\\
	\center
	\includegraphics[width=0.85\linewidth]{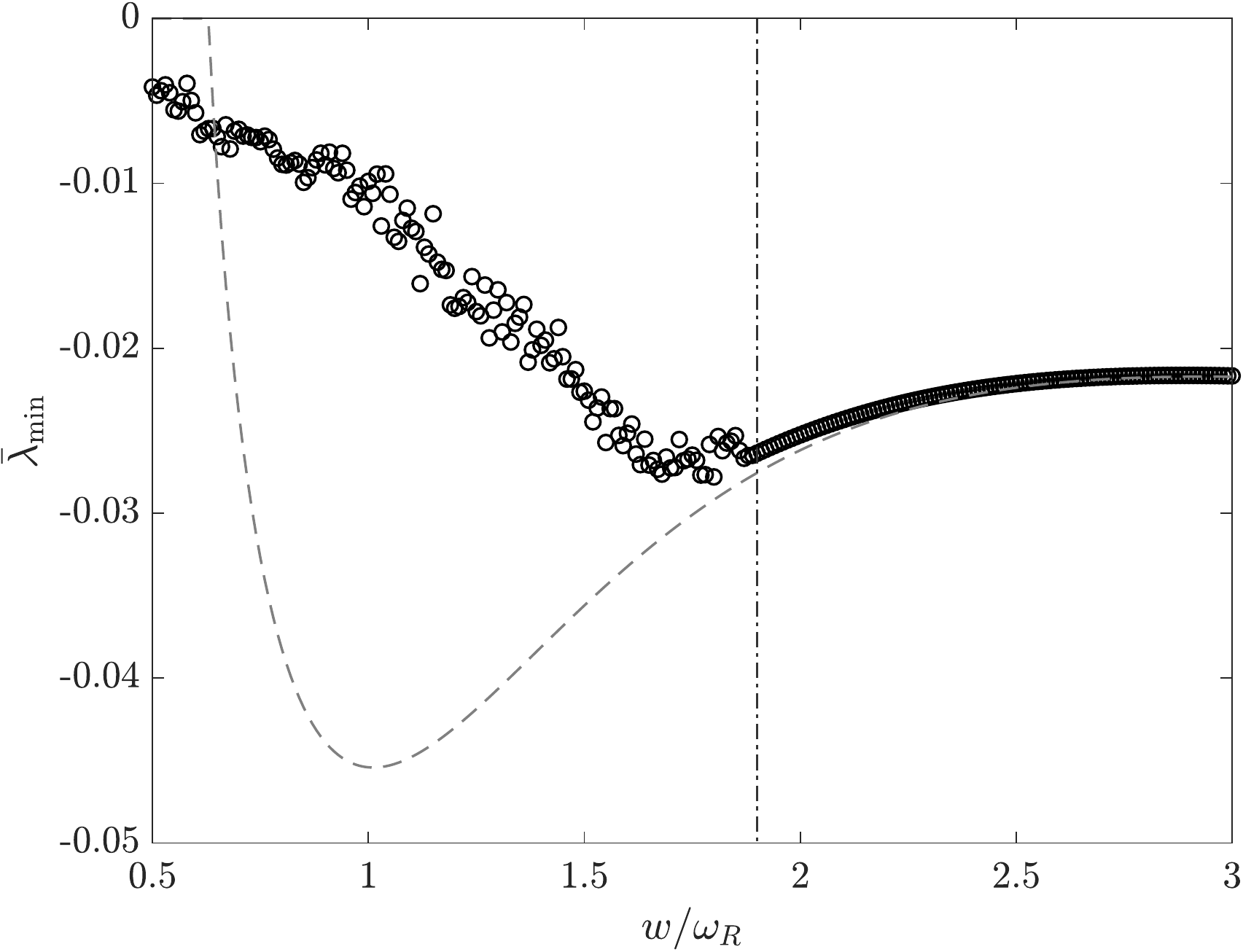}\\
	\flushleft(b)\\
	\center
	\includegraphics[width=0.85\linewidth]{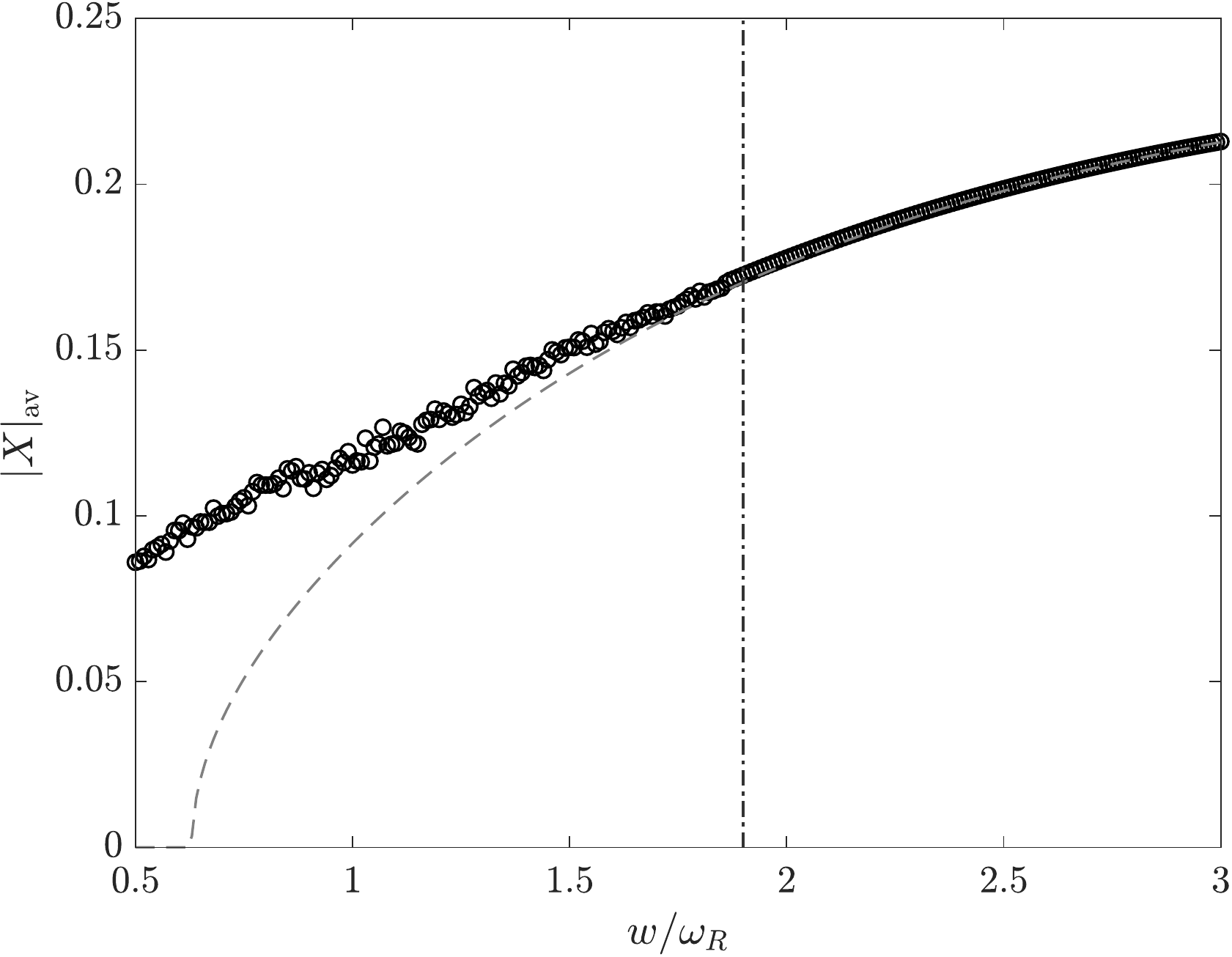}\\
	\caption{(a) The time-averaged minimum eigenvalue $\bar{\lambda}_{\mathrm{min}}$ calculated using Eq.~\eqref{tamineig} (black circles) with $t_{\mathrm{end}}=10^4\omega_R^{-1}$ as a function of $w$ in units of $\omega_R$. The gray, dashed line corresponds to the value of $\lambda_{\mathrm{min}}$ calculated using the iterative method explained in section ``{\bf Asymptotic state}''. (b) The time-averaged mean-field order parameter $|X|_{\mathrm{av}}$ (black circles) with $t_{\mathrm{end}}=10^4\omega_R^{-1}$ and the asymptotic result (gray, dashed line) as a function of $w$ in units of $\omega_R$. The vertical dashed lines in subplot (a) and (b) show the critical pumping strength $w_c(\Lambda)$. The equations are integrated in the momentum interval $[-15\hbar k, 15\hbar k]$ for $\Lambda=15\omega_R$ and $\Delta=\kappa/2$. \label{Fig:S1}}
\end{figure}

In the main article we show that the system relaxes to an incoherent state for $\Lambda<\Lambda_c$. In that case, at steady state, all particles are in the excited state and therefore the system has no entanglement between internal and external degrees of freedom. However, if the system is in the coherent phase, for $\Lambda >\Lambda_c$, we observe entanglement between internal and external degrees. This claim can be verified by an analysis of the partial transpose $\hat{\rho}_{\mathrm{st}}^{\mathrm{PPT}}$ of the stationary state $\hat{\rho}_{\mathrm{st}}$ \cite{Horodecki:2009}. The matrix $\hat{\rho}_{\mathrm{st}}^{\mathrm{PPT}}$ is calculated from $\hat{\rho}_{\mathrm{st}}$ by applying the transpose on the internal degrees of freedom only. In the case where internal and external degrees of freedom are not entangled $\hat{\rho}_{\mathrm{st}}^{\mathrm{PPT}}$ is a positive matrix. On the other hand if $\hat{\rho}_{\mathrm{st}}^{\mathrm{PPT}}$ is not positive we know that internal and external degrees must be entangled.
To check whether the internal and external degrees of freedom are entangled in the coherent phase we calculate the minimum eigenvalue
\begin{align}
	\lambda_{\mathrm{min}}=\min\{\lambda|\lambda\text{ is eigenvalue of }\hat{\rho}_{\mathrm{st}}^{\mathrm{PPT}}\}.\label{lambdamin}
\end{align}
If this eigenvalue is negative we know that the system is entangled. The numerical calculated values for $\lambda_{\mathrm{min}}$ are shown in Fig.~4(c).
As expected we observe that there is no entanglement for $\Lambda<\Lambda_c$. For $\Lambda >\Lambda_c$ we observe a negative $\mathrm{\lambda}_{\mathrm{min}}$, demonstrating that in this region, the stationary state is an entangled state.

For the transition from chaotic to coherent the analysis needs to be adjusted. While in the coherent phase we can apply the method that we explained above we cannot use this method in the chaotic phase since there is no stable stationary state. Therefore we introduce a time-averaged minimum eigenvalue \begin{align}
	\bar{\lambda}_{\mathrm{min}}=\frac{1}{t_{\mathrm{end}}}\int_0^{t_{\mathrm{end}}}\lambda_{\mathrm{min}}(t)dt.\label{tamineig}
\end{align} 
To calculate $\bar{\lambda}_{\mathrm{min}}$ we need to calculate $\mathrm{\lambda}_{\mathrm{min}}(t)$ as a function of time
\begin{align}
	\lambda_{\mathrm{min}}(t)=\min\{\lambda|\lambda\text{ is eigenvalue of }\hat{\rho}(t)^{\mathrm{PPT}}\}.\label{lambdamintime}
\end{align}
We plot $\bar{\lambda}_{\mathrm{min}}$ in Fig.~S\ref{Fig:S1}(a) where we observe that $\bar{\lambda}_{\mathrm{min}}<~0$ for the shown interval of $w$. The calculated value of the minimal eigenvalue from the iterative method is shown as the gray, dashed line. While the results of both methods agree in the coherent phase $w>w_c$ we observe large discrepancies in the chaotic phase. 

For completeness we also report the time-averaged mean-field order parameter 
\begin{align}
	|X|_{\mathrm{av}}=\frac{1}{t_{\mathrm{end}}}\int_0^{t_{\mathrm{end}}}|X(t)|dt\label{Xav}
\end{align}
in Fig.~S\ref{Fig:S1}(b). The time-averaged mean-field order parameter $|X|_{\mathrm{av}}$ and the asymptotic result agree in the coherent phase while $|X|_{\mathrm{av}}$ is larger in the chaotic phase. 
Notice that the discrepancies in the chaotic phase are expected since here the asymptotic state is not a stable state and the description in terms of a single stationary state fails.

\end{document}